\documentclass[aps,prb,twocolumn,showpacs]{revtex4-1}

\usepackage{graphicx}
\usepackage{multirow}
\usepackage{fixltx2e}

\usepackage{amsmath}
\usepackage[normalem]{ulem}
\usepackage{color}

\begin{document}

\title{Localization versus inhomogeneous superfluidity: 
Submonolayer $^4$He on fluorographene, hexagonal boron nitride, 
and graphene}

\author{Saverio Moroni}
\affiliation{CNR-IOM  Democritos and SISSA, Via Bonomea 265, 34136 Trieste, Italy}

\author{Francesco Ancilotto}
\affiliation{Dipartimento di Fisica e Astronomia ’Galileo Galilei’ and CNISM, Universit\`a di Padova, via Marzolo 8, 35122 Padova, Italy}
\affiliation{CNR-IOM  Democritos, Via Bonomea 265, 34136  Trieste,  Italy}

\author{Pier Luigi Silvestrelli}
\affiliation{Dipartimento di Fisica e Astronomia ’Galileo Galilei’ and CNISM, Universit\`a di Padova, via Marzolo 8, 35122 Padova, Italy}
\affiliation{CNR-IOM  Democritos, Via Bonomea 265, 34136  Trieste,  Italy}

\author{Luciano Reatto}
\affiliation{Dipartimento di Fisica, Universit\`a degli Studi di Milano,via Celoria 16, 20133 Milano, Italy}

\begin{abstract}
We study a sub monolayer $^4$He 
adsorbed on fluorographene (GF) and on hexagonal boron nitride (hBN) at low 
coverage. The adsorption potentials have been computed {\it ab initio} with a 
suitable density functional theory including dispersion forces. The
properties of the adsorbed $^4$He atoms have been computed 
at finite temperature with path integral 
Monte Carlo 
and at $T=0$~K with variational path integral.
From both methods we find that the lowest energy state 
of $^4$He on GF is a superfluid. Due to the very large corrugation of the 
adsorption potential this superfluid has a very strong spatial anisotropy, 
the ratio between the largest and smallest areal density being about 
6, the superfluid fraction at the lowest $T$ is
about 55\%, and the temperature of the transition to the normal state is 
in the range 0.5-1 K. Thus, GF offers a platform for studying the properties 
of a strongly interacting highly anisotropic bosonic superfluid. 
At a larger coverage $^4$He has a transition to an ordered commensurate 
state with occupation of 1/6 of the adsorption sites. 
This phase is stable up to a transition temperature
located between 0.5 and 1~K. The system has a triangular order similar to
that of $^4$He on graphite but each $^4$He atom is not confined to a 
single adsorption site and the atom visits also the nearest neighboring 
sites giving rise to a novel three--lobed density distribution. The lowest 
energy 
state of $^4$He on hBN is an ordered commensurate
state with occupation of 1/3 of the adsorption sites and triangular symmetry. 
A disordered state is present at lower coverage as a metastable state. 
In the presence of an electric field the corrugation of the adsorption potential 
is slightly increased but up to a magnitude of 1~V/\AA~ the effect is small 
and does not change the stability of the phases of $^4$He on GF and hBN. 
We have verified that also in the case of graphene such electric field 
does not modify the stability of the commensurate 
$\sqrt{3}\times\sqrt{3}R30^\circ$ phase.
\end{abstract} 

\maketitle

\section{Introduction}
\label{sec_intro}
Bosons moving in a periodic external potential $V_{\rm ext}({\bf r})$ can 
be found in two different states: depending on the amplitude of 
the modulation of the external potential either the bosons are 
localized at the minima of  $V_{\rm ext}({\bf r})$ or the bosons are 
delocalized and superfluid at low temperature. For very strong interboson 
interaction an incommensurate solid can be present. 
These two regimes, localized or 
delocalized, have been achieved with cold bosonic atoms moving in 
the periodic potential generated by optical standing waves 
\cite{greiner_2002}. 
Also a submonolayer film of light bosons like $^4$He atoms adsorbed 
on a crystalline substrate is expected to show one of these two 
regimes depending on the character of the adsorption potential. 
In practice so far the only substrate that can approach such ideal 
situation is graphite because this material can be obtained with 
an extended almost perfect surface on which one can study the 
properties of the adsorbed species \cite{brunch_2007}. 
It turns out that the 
He-graphite adsorption potential is characterized by a corrugation 
that is large enough so that the He atoms are localized around the 
preferential adsorption sites and in fact experiment 
\cite{brunch_2007} and 
theory \cite{pierce_1999,manousakis_recente}
agree that the ground state of a monolayer $^4$He is 
an ordered commensurate state in which the $^4$He atoms occupy one 
third of the adsorption sites so the adatoms have a crystalline 
triangular symmetry, the so-called $\sqrt{3}\times\sqrt{3}R30^\circ$
phase. 
Therefore $^4$He on graphite is nonsuperfluid, superfluidity 
appearing only when at least two layers are present 
\cite{reppy,nyeki_2017}.

On theoretical basis a similar behavior is expected for the 
adsorption of $^4$He on graphene (G) because its adsorption potential 
turns out to be very similar to that of graphite. Other 
substrates commonly used in adsorption studies are intrinsically 
disordered like that of a glass or are in any case too disordered 
to be relevant on this issue. It would be of great interest to 
find other materials with extended crystalline surfaces with a 
corrugation of the adsorption potential smaller than that of 
graphite because this would give the possibility of a new 
superfluid state of strongly interacting particles that will be 
spatially anisotropic due to the influence of the substrate 
potential. It has been of interest that recent theoretical 
studies \cite{nava_2012,reatto_2013}
found that for two derivatives of graphene, 
fluorographene (GF) and graphane (GH), the adsorption potential 
is very different from that of graphene and of graphite and it 
turns out that the $^4$He atoms are delocalized and the ground 
state of monolayer $^4$He on GF and on GH was claimed to be
superfluid. That study was based on a semi empirical adsorption 
potential. These results have been corroborated by a more recent 
study \cite{silvestrelli_2019}
in which some of us have developed an adsorption 
potential based on ab intio methods: even with such adsorption 
potential it was found that the $^4$He atoms on GF are delocalized 
and the ground state of submonolayer $^4$He is a superfluid. 
Interestingly, that work \cite{silvestrelli_2019} has shown that
also the monolayer of $^4$He on hexagonal boron nitride (hBN) 
turned out to be a superfluid.

Studies of Refs. \cite{nava_2012,reatto_2013,silvestrelli_2019}
are based on state of the art many 
body computations that should be able to provide exact results 
for the system. However quantum simulations of particles in a 
highly structured potential can be tricky due to the multiple 
energy scales that are present, and we decided to perform a new 
investigation of the adsorption of $^4$He atoms on GF, hBN, and G. 
In the present study,
we have also derived 
the adsorption potential in presence of an external electric 
potential, that is a possible way to alter the corrugation of 
the adsorption potential. 

We have studied $^4$He on GF and hBN at $T=0$~K with 
the variational path integral (VPI) method \cite{ceperley_1995}, 
also known as path integral ground state (PIGS) \cite{sarsa_2000},
and at finite temperature with 
path integral Monte Carlo (PIMC) method \cite{ceperley_1995}.
In a VPI computation 
the quantum state is obtained by projecting an assumed initial 
state with the imaginary time evolution operator. For large enough 
propagation time $\beta$ one gets an unbiased sampling 
\cite{rossi_2009}
of the 
properties of the exact ground state of the system if the initial 
state is not orthogonal to the ground state. How large $\beta$ has to 
be must be found empirically in terms of convergence. 
Our evidence is that the results in 
\cite{nava_2012,reatto_2013,silvestrelli_2019}
were not fully converged. We are 
confident that the present VPI results are at convergence also 
because they are in agreement, as expected, with those of 
PIMC at low temperature.

In the case of $^4$He on GF we confirm the main result of the 
previous studies \cite{nava_2012,silvestrelli_2019}: 
The ground state of $^4$He on GF is a 
superfluid that has a strong spatial modulation. At variance with 
the result of Refs.~\onlinecite{nava_2012,silvestrelli_2019},
we find that at coverage larger 
than that of the equilibrium state there is a first order
phase transition to a commensurate state with triangular symmetry at a 
coverage corresponding to the occupation of 1/6 of the adsorption 
sites. 

On the other hand, in the case of $^4$He on hBN, we do not confirm the 
previous result of the existence of a superfluid state and our 
results show that, in the ground state and at low temperature, 
$^4$He atoms on hBN form a nonsuperfluid commensurate state very 
similar to that of $^4$He on graphite.

For all the considered substrates we also find that the change of the adsorption 
potential due to the application of an external electric field 
$E$ is not large enough to alter in a significant way the properties 
for $E=0$, at least for the strength of $E$ allowed by our 
approximations (see below).

The contents of the paper are as follows. In Section \ref{sec_potentials}
we present the results for 
the adsorption potential of He on GF, hBN, and G with and without 
an external electric field. The quantum simulations of our study 
are described in Section \ref{sec_qmc}.
A summary and our conclusion are contained in Section \ref{sec_summary}.
Technical details of the quantum simulations are given in the Appendix.

\section{Potentials}
\label{sec_potentials}
Monolayer $^4$He films on GF and GH have been 
proposed recently as novel superfluid systems characterized by 
strong in-plane anisotropies.
This remarkable prediction \cite{nava_2012,Nav12,Nav13,reatto_2013} 
was based on quantum simulations where an essential ingredient
is an accurate description of the
interaction between He atom and the substrates.
Specifically,
the He-substrate interaction potential 
was modeled using a traditional
semiempirical approach\cite{Sto80}, where the potential energy of a
single He atom near the surface is written 
as a sum of pair potential interactions made of 
a repulsive part proportional to the local
electron density (Hartree-Fock repulsion),
and an attractive part, in the form
of a sum of damped He atom van der Waals (VdW)
interactions and polarization interaction due to
the surface electric field \cite{nava_2012,Nav12,Nav13}.

These effective potentials are known to be affected by 
quite large uncertainties in the empirical coefficients
used to model the interaction.
The importance of a precise knowledge of the
adatom-surface interaction potentials
to make quantitative prediction on the adsorption 
properties of surfaces cannot be overlooked: the wetting
properties of rare-gas atoms on solid surfaces, for instance, are
known to strongly depend on the strength and
corrugation of the adatom-substrate potential.

For this reason, we decided to investigate from first principles
the interaction of He atoms with GF and hBN.
using state-of-the-art Density Functional Theory (DFT) functionals
specifically designed to describe the weak VdW interactions,
with the goal of providing a more accurate description
of the interaction of He atoms with these surfaces.
Recent applications of vdW-corrected DFT schemes to the
problem of atoms/molecules-surface interactions
have proven the accuracy of such methods
in the calculation of both adsorption distances and adsorption
energies, as well as the high degree of its reliability
across a wide range of adsorbates.
In particular, we have computed 
the He atom adsorption energies on different surface sites
and the potential energy corrugations along the plane,
which are the most crucial ingredients for accurate quantum simulations
of the adsorption of $^4$He.
Our calculations have been performed
with the Quantum-ESPRESSO {\it ab initio} package\cite{ESPRESSO}.
A single He atom per supercell is considered and we model the substrates
adopting periodically repeated orthorhombic supercells,
with a $4 \times 2$ structure, in the case of G with 64 C atoms,
in the case of GF with 32 C atoms plus as many F atoms, while
in the case of hBN the substrate 
is formed by 16 B and 16 N atoms. The
lattice constants correspond to the equilibrium state of the substrates. 
Repeated slabs were separated along the direction orthogonal to the 
surface by a vacuum region
of about 24 \AA\ to avoid significant spurious interactions due to periodic
replicas. The Brillouin Zone has been sampled using a
$2\times2\times1$ $k$-point mesh.
Electron-ion interactions were described using ultrasoft
pseudopotentials and the wavefunctions were expanded in a plane-wave basis 
set with an energy cutoff of 51 Ry.

The calculations have been performed by adopting the  
rVV10\cite{Sabatini} DFT functional (this is the revised, more efficient 
version of the original VV10 scheme\cite{Vydrov}), where
vdW effects are included by introducing an explicitly nonlocal correlation 
functional. 
rVV10 has been found to perform well in many systems and processes 
where vdW effects are relevant, including several adsorption 
processes\cite{Sabatini,psil15,psil16}.
This DFT functional is able to well reproduce the reference
structural data of GF, hBN, and G including
the ``buckling displacement'' in GF\cite{nava_2012,Nav12,Nav13}.  

We have also investigated the effect of the application of an external 
uniform electric field. Since in supercell calculations periodic boundary 
conditions are imposed on the electrostatic potential, an external 
electric field is simulated by adding a sawlike potential to the bare 
ionic potential\cite{Neugebauer}, also including a dipole correction, 
according to the recipe of Bengtsson\cite{Bengtsson}.
This represents the proper way to simulate an external electric field in
surface calculations with a slab geometry, provided the electrostatic 
potential discontinuity falls in the middle of the vacuum 
region\cite{Neugebauer,Bengtsson,Meyer}.
A positive value means that the external electric field points away
from the surface in the positive $z$ direction (the adsorbed He atom
is located in the positive $z$ region).
The additional external potential also leads to changes in the total energy
of the system and the Hellmann-Feynman forces\cite{Meyer}.

An upper limit for the amplitude of the external electric field, which
can be considered in the simulations, exists that is determined by the
thickness of the vacuum region. In fact a kind of ``quantum well'' is
formed in the vacuum region\cite{Meyer}; if the electrostatic
potential of this quantum well drops below the Fermi level, it can
become populated by the transfer of electrons from the slab region:
the threshold for the occurrence of this unwanted, artificial behavior
depends on both the width of the vacuum region and the strength of
the electric field\cite{Meyer}.
In our applications this upper limit turns out to be around 1.0 V/\AA, 
which is a value of the same order (or even larger) than the maximum
electric field achievable in actual experiments (for instance, in
electrowetting applications\cite{Ostrowski}). 

\begin{figure}[h]
\includegraphics[width=8cm]{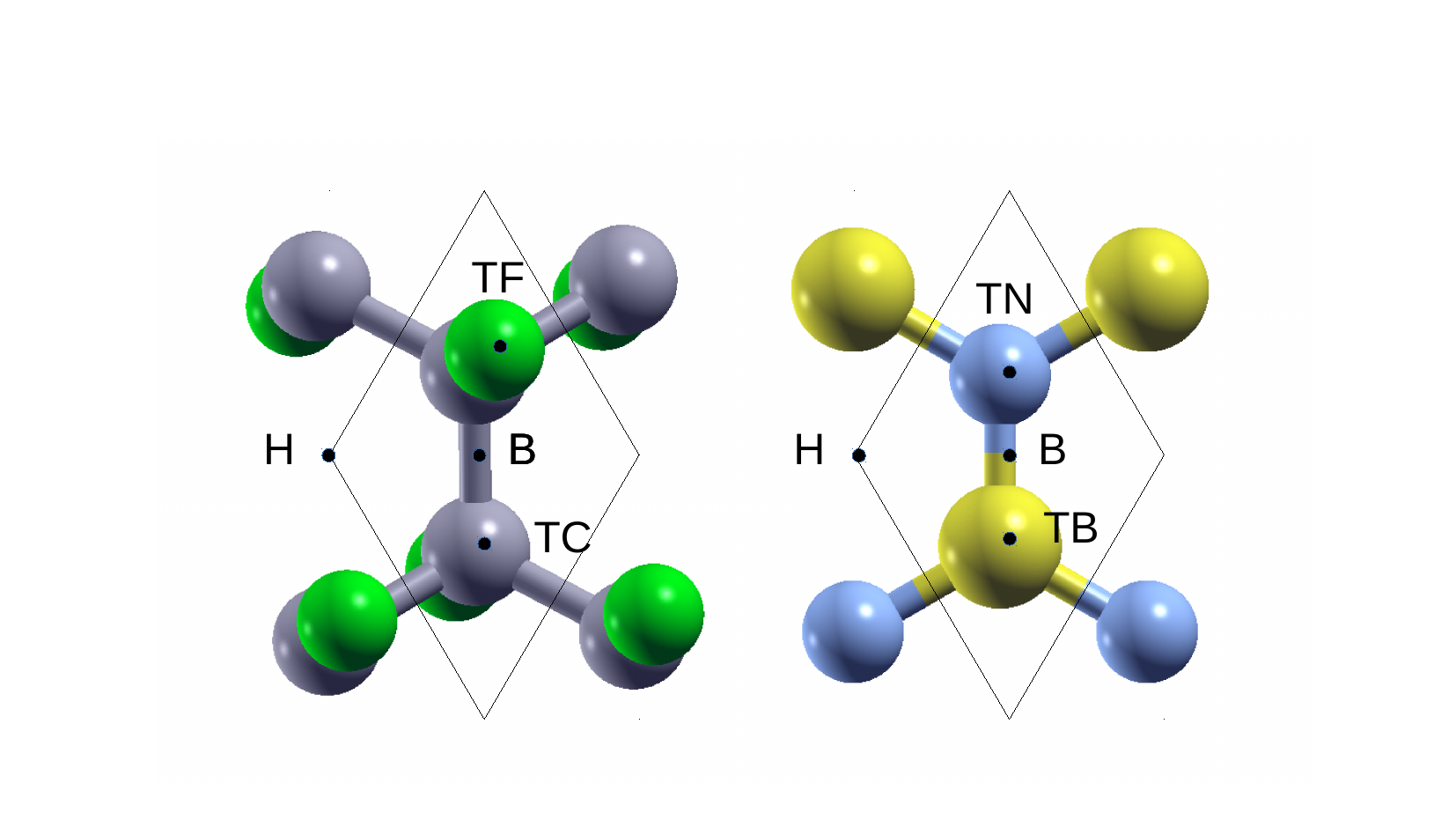}
\caption{Sites chosen for the calculation of the He-GF (left panel) and He-hBN (right panel) potentials. GF has a three-layer
structure: The central layer is occupied by the carbon atoms as in a graphene 
sheet except for a small vertical offset between the two triangular 
sublattices, the upper plane and the lower one contain the fluorine atoms. 
In the left panel C atoms are shown as grey circles and F atoms as 
green circles. hBN is a single layer structure isomorphous to graphene. 
In the right panel B atoms are shown as yellow circles and N atoms as 
blue circles.
\label{fig:sites_gf}
}
\end{figure}

We computed the He-substrate interaction (with and without electric field)
for a selected set of nonequivalent 
sites in the 
primitive surface unit cell. 
In particular, for He on GF, we chose the following points:
$H$ (hollow site), $TC$ (site on top of C atom), $TF$ (site on top of F 
atom), $B$ (bridge site between C and F);
for He on hBN we chose:
$H$ (hollow site), $TB$ (site on top of B atom), $TN$ (site on top of N 
atom), $B$ (bridge site between B and N); 
finally, for He on G, we considered the points: $H$ (hollow site), 
$TC$ (site on top of C atom), and $B$ (bridge site between two adjacent C atoms). 
The sites for GF and
hBN are shown in Fig.~\ref{fig:sites_gf}.
 
\begin{figure}[h]
\includegraphics[width=8cm]{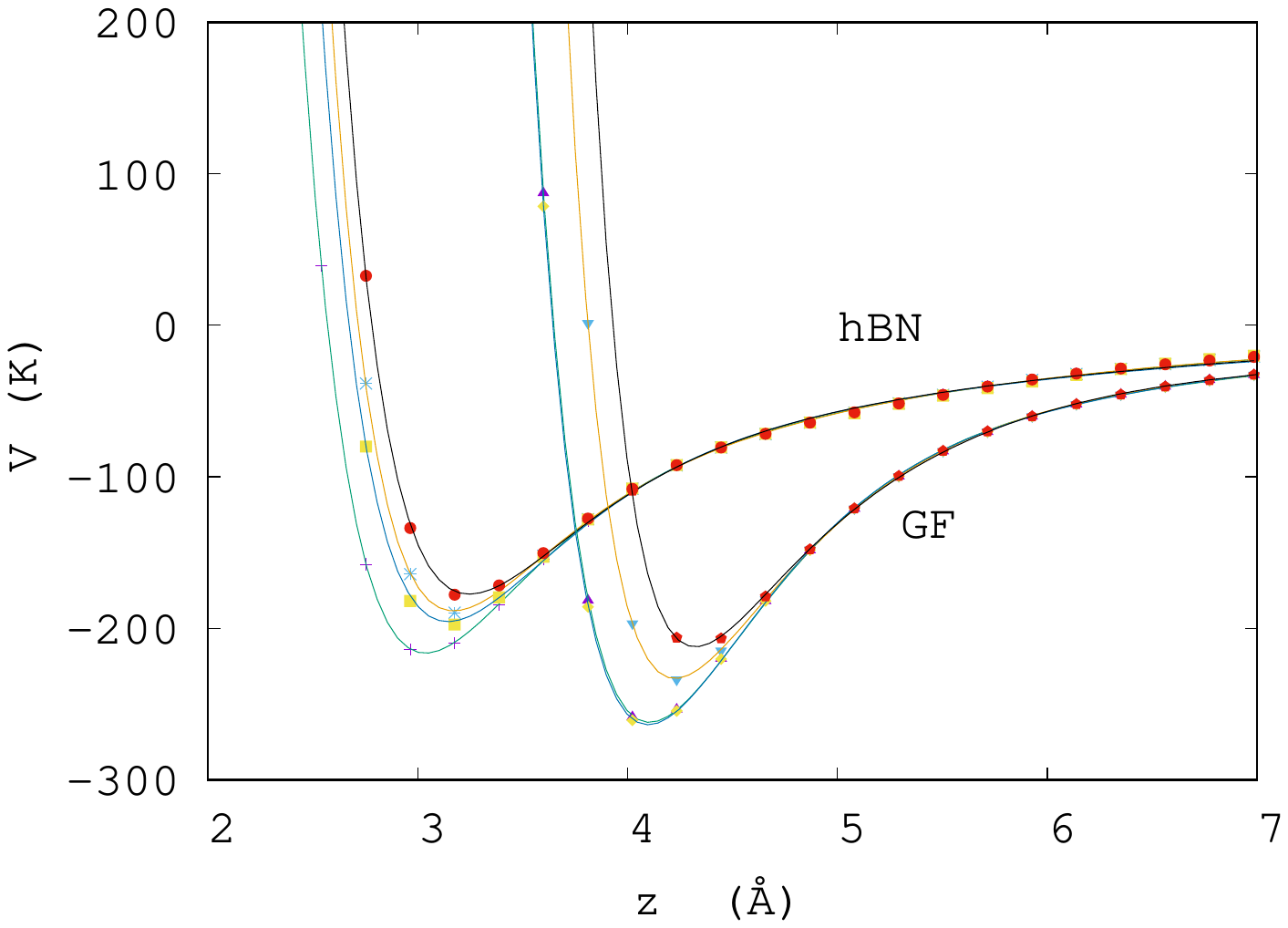}
\caption{
Calculated He-hBN and He-GF interaction potentials along the $z$ 
direction above selected sites.
GF (from top to bottom): $TF$, $B$, $H$, $TC$;
hBN (from top to bottom): $TN$, $B$, $TB$, $H$.
The points show the {\it ab initio} values, the lines
are the results of 8-parameter curve fitting with the form
$
\sum _{i=1}^{2} a_i \, exp(-b_i z)-\sum _{i=1}^{4}c_i/z^{2i+2},
$
whose parameters are given in Table~\ref{tab:table1}.
\label{fig_potentials}
}
\end{figure}

Besides the lowest-energy configurations for a given investigated 
adsorption site, we have also computed the dependence upon 
the normal coordinate $z$ of the He-substrate interaction potentials 
above those sites, shown in Fig.~\ref{fig_potentials}. 
Our goal was to provide a 
reliable three-dimensional He-substrate potential function 
to be used for numerical simulations
based on Quantum Monte Carlo methods, that will be discussed in the 
following.
\begin{table*}[!]
\caption{
Best-fit parameters for the $z$ dependence of the He-substrate potentials
shown in Fig.~\ref{fig_potentials}.
}
\begin{center}
\label{tab:table1}
\begin{tabular}{cccccccccc} 
\hline
\hline    
$ $ & $ $ & $a_1$ & $b_1$ & $a_2$ & $b_2$ & $c_1$ & $c_2$ & $c_3$ & $c_4$ \\
\hline
GF: & $TF$ & $2.53645\times 10^7$ & $1.55209$ & $1.07222\times 10^6$ & $0.842514$ & $-4.49616\times 10^6$ & $7.37512\times 10^8$ & $-5.63025\times 10^9$ & $1.19823\times 10^{10}$ \\
    & $B$ &  $-4.25772\times 10^8$ & $2.63907$ & $-1.0177\times 10^7$ & $1.61871$ & $-339981$ & $7.38605\times 10^7$ & $-3.58557\times 10^9$ & $1.1976\times 10^{10}$ \\
    & $H$ & $-2.73175\times 10^6$ & $1.91148$ & $-2.77692\times 10^6$ & $1.91101$ & $42978.4$ & $8.18279\times 10^6$ & $-3.9442\times 10^8$ & $1.54338\times 10^9$ \\
    & $TC$ & $526386$ & $0.975559$ & $20543.4$ & $0.389085$ & $4.68048\times 10^6$ & $8.32948\times 10^6$ & $-4.34654\times 10^8$ & $1.61137\times 10^9$ \\
\hline
hBN: & $TN$ & $-244960$ & $1.94545$ & $-231170$ & $1.94594$ & $93994$ & $-1.97338\times 10^6$ & $1.6644\times 10^6$ & $2.86811\times 10^6$ \\
     & $B$ & $2.15693\times 10^7$ & $3.62009$ & $2.03143\times 10^7$ & $3.61956$ & $87334$ & $-2.0872\times 10^6$ & $2.18302\times 10^7$ & $-3.33307\times 10^7$ \\
     & $TB$ & $-208875$ & $1.90869$ & $-218649$ & $1.90819$ & $99505.8$ & $-2.25015\times 10^6$ & $4.40446\times 10^6$ & $-1.32199\times 10^6$ \\
     & $H$ & $1.75506\times 10^7$ & $3.56209$ & $1.7902\times 10^7$ & $3.56201$ & $89581.7$ & $-2.20804\times 10^6$ & $2.3581\times 10^7$ & $-3.80861\times 10^7$ \\
\hline
\hline
\end{tabular}
\end{center}
\end{table*}

We approximate such potential by using a truncated Fourier
expansion over the first three stars of the two-dimensional reciprocal 
lattice associated with a triangular lattice with a two-atom basis 
(two C atoms in the case of G, one C and one F atom in the case of GF, 
one B and one N atom in the case of hBN). 
The Fourier components can be obtained from the calculated $z$ dependence 
of the various symmetry sites described above.\cite{nota}

\begin{table}
\caption{Binding energy in the lowest-energy configuration
(with and without external electric field),
$E_b$, distance $d$ of He from the reference plane (defined by
averaging over the $z$ coordinates of the C atoms for G and
GF, and of the B and N atoms for hBN), maximum corrugation,
$\Delta_{\rm{max}}$, minimum intersite energy-barrier, $\Delta_{\rm{min}}$
(see text for the definitions), for He-G, He-GF, and for He-hBN, using the
rVV10 DFT functional.}
\begin{tabular}{|c|l|r|c|r|r|}
\hline\noalign{\smallskip}
 system & electric field (V/\rm{\AA}) & $E_b$(K) & $d(\rm{\AA})$ &$\Delta_{\rm max}$(K) & $\Delta_{\rm min}$(K) \\ 
\hline
 He-GF  & 0.0            &   -261   & 4.10 &     54   &       11   \\ 
 He-GF  & 1.0            &   -278   & 4.10 &     55   &       10   \\ 
\hline
 He-hBN & 0.0            &   -214   & 2.96 &     36   &       16   \\ 
 He-hBN & 1.0            &   -235   & 2.96 &     43   &       19   \\ 
\hline
 He-G   & 0.0            &   -298   & 2.96 &     50   &       47   \\ 
 He-G   & 0.6            &   -290   & 2.96 &     47   &       44   \\ 
\noalign{\smallskip}\hline
\end{tabular}
\label{table-energy}
\end{table}
Our numerical results for the adsorption of He on the different substrates
we have considered are summarized in Table~\ref{table-energy}.
In particular, we report the distance $d$
of He from the substrate and the binding energy $E_b$ of the
the lowest-energy configuration.
This configuration is found above the hollow site $H$ for hBN and G, and 
above the $TC$ site for GF (see Fig.~\ref{fig_potentials}). 
For GF a nearly isoenergetic configuration is found also above
the hollow site \cite{nava_2012,silvestrelli_2019}.
Therefore on GF two essentially equivalent kinds of adsorption sites are 
present and the overall number of adsorption sites is double that
on hBN and G.
In Table~\ref{table-energy} we also report two other energetic parameters: 
the "maximum
corrugation", $\Delta_{\rm{max}}$, defined as the difference between
the binding energy of He
on top of C, F, and N atom
(which represents the less-favored configuration for
G, GF, and hBN, respectively) and the binding energy of the
lowest-energy configuration, and the
"minimum intersite energy barrier", $\Delta_{\rm{min}}$,
which is given by the minimum energy barrier that the He atom must
overcome to be displaced from an optimal adsorption site to another,
namely from $H$ to $H$ for hBN and G and from $H$ to $TC$ for
GF.
This latter quantity has been evaluated by monitoring
the binding energy corresponding to a reaction path generated by
constraining the planar $x$, $y$ coordinates of the He atom and optimizing
the vertical $z$ coordinate only.
In the case of hBN, $\Delta_{\rm{min}}$ ($\Delta_{\rm{max}}$)
corresponds to the difference between the binding energy of He on top of
the B(N) atom, $TB$($TN)$ site, and the binding energy of the
lowest-energy configuration, $H$.

For the case of no external electric field, the results for the corrugation
of the adsorption potential agree with those
reported in previous studies:\cite{nava_2012,silvestrelli_2019,nuova_nota}
the most striking difference between the
case of He-G and of He-GF is that in He-G $\Delta_{\rm{max}}$ and
$\Delta_{\rm{min}}$
are comparable (the difference is about 7\%) while, on the contrary, 
in He-GF $\Delta_{\rm{min}}$ is
smaller than $\Delta_{\rm{max}}$ by a factor 5. This
confirms that the adsorption potential of He on GF is characterized
by narrow "canyons" between adsorption sites, with a much larger anisotropy
in the corrugation and a relatively low energy barrier compared to G.
A large difference between $\Delta_{\rm{min}}$ and $\Delta_{\rm{max}}$
(about a factor 2)
is found also for hBN
so that the corrugation is larger than in the case of G.
Another difference between hBN and G is that each adsorption site in hBN 
is surrounded by three saddle points and not six as in the case of G and graphite. 
It should be noticed that the adsorption energy of He on GF found with the 
semiempirical approach in Ref.~\onlinecite{nava_2012} appears to be 
strongly overestimated, we 
find that this energy is about 10\% smaller than that on G and not much 
larger as reported in Ref.~\onlinecite{nava_2012}.

When an external, uniform electric field is applied
(with the maximum strength of $E$ allowed by our approximations,
see discussion above), there is a change of the adsorption potential,
leading
to moderate quantitative changes in the quantities reported in 
Table~\ref{table-energy},
which,
however, are no so significant to alter the basic properties found
at vanishing electric field.

\section{Quantum simulations}
\label{sec_qmc}
We use the rVV10 potentials of Sec.~\ref{sec_potentials} to calculate 
unbiased equilibrium thermal averages and ground-state properties of $^4$He 
adsorbed on GF, hBN and G with quantum Monte Carlo simulations. 
For $N$ He atoms the Hamiltonian is
\begin{equation}
H=-\frac{\hbar^2}{2m}\nabla^2+\sum_{i=1}^N
+v_1({\bf r}_i)+\sum_{i<j}v_2(r_{ij}),
\label{eq:hamiltonian}
\end{equation}
where $v_1$ is the appropriate He-substrate one-body potential
with ${\bf r}_i$ the position of the $i$th $^4$He atom,
and $v_2$ is the Aziz HFDHE2 He-He pair potential\cite{aziz}
with $r_{ij}=|{\bf r}_i-{\bf r}_j|$. 
The substrate is placed at $z=0$ with the $^4$He atoms
in the positive $z$ semispace, and periodic boundary conditions are 
applied in the $x$ and $y$ directions.

The finite temperature simulations are performed using the PIMC 
method \cite{ceperley_1995} with the worm algorithm \cite{boninsegni_2006}. 
In this approach the matrix element
$\langle R|e^{-\beta H}|PR\rangle$, where $\beta$ is the inverse
temperature, $R=\{{\bf r}_1,\ldots,{\bf r}_N\}$, and
$PR$ is a permutation thereof, is represented as a real-space convolution of
$M$ high-temperature density matrices $\langle R'|e^{-\tau H}|R''\rangle$, 
where $\tau=\beta/M$, which are in turn approximated by suitable closed-form
expressions. The results are unbiased in the limit $\tau\to 0$.
We use the primitive approximation\cite{ceperley_1995} with 
$\tau=0.002~{\rm K}^{-1}$. In the Appendix we show that the
finite-$\tau$ systematic error is negligible for our purposes.
We work in the grand canonical ensemble, i.e. at fixed volume $V$, 
temperature $T$ and chemical potential $\mu$. 

The ground state simulations are performed with a projection Monte Carlo
method known as VPI \cite{ceperley_1995} or 
PIGS \cite{sarsa_2000}. This approach uses the same real-space convolution 
representation of the operator $e^{-\beta H}$ mentioned above for PIMC to 
calculate expectation values on the quantum state 
$\Psi_\beta=e^{-\beta H}\Psi$, where the trial function $\Psi$ is a 
closed-form approximation to the ground state. 
We use a trial function of the form
\begin{equation}
\Psi(R)=\exp\left[-\sum_i u_1(z_i)-\sum_{i<j}u_2(r_{ij})\right],
\label{eq:psi}
\end{equation}
where the one-body He-substrate pseudopotential $u_1$ and the two-body
He-He pseudopotential $u_2$ are parametrized with six variational
parameters each and optimized by energy minimization using the stochastic
reconfiguration method\cite{rocca}.
For $\beta\to\infty$ 
the state $\Psi_\beta$ approaches the exact ground state. Thus, in 
addition to the finite--$\tau$ bias, one has to control the error
incurred using a necessarily finite value of $\beta$. We use 
$\beta=2~{\rm K}^{-1}$; examples of convergence in $\beta$ are
shown in the Appendix.

We use either ``small'' or ``large'' simulation cells,
respectively containing 180 or 720 adsorption sites for GF,
and half that many sites for hBN and G.
For GF we will also 
present a few results for a larger cell with 2880 sites.

\subsection{GF}

\begin{figure}[h]
\includegraphics[width=8cm]{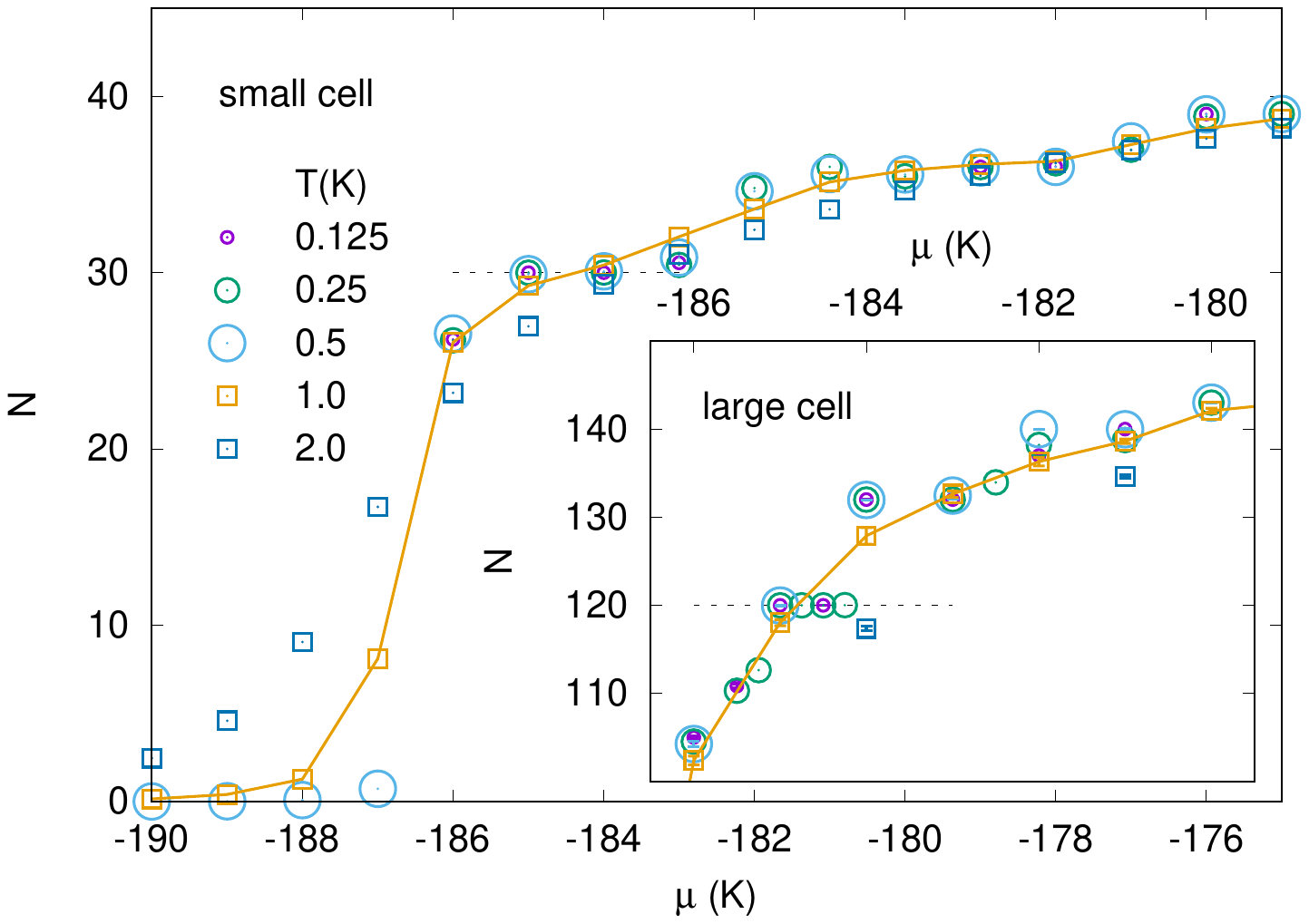}
\caption{
$^4$He adsorbed on GF: number of atoms as a function of the chemical 
potential calculated for various temperatures, as indicated, 
for the small cell (main figure) and the large cell (inset). 
The dashed lines indicate the numbers at
filling 1/6.
\label{fig_gf_nofmu_s}
}
\end{figure}

The initial configuration for our PIMC simulations in the grand 
canonical ensemble is the empty cell. After equilibration, the
number of $^4$He atoms fluctuates around a stationary average value $N$ which
depends on the input chemical potential $\mu$. Figure~\ref{fig_gf_nofmu_s} 
shows a plot of $N(\mu)$ at various temperatures.
For the small cell (main figure) the data for $T=2$~K follow a smooth curve, 
corresponding to a fluid phase extending over the whole range of $\mu$
shown in the plot.
Upon lowering the temperature
below $T=1$~K, a flat region at $N(\mu)=30$ develops 
between $\mu=-185$ and -184~K. Here the $^4$He atoms form a
commensurate solid phase with areal density 
$\rho=0.0574$~\AA$^{-2}$, at 1/6 of the adsorption sites.
A fluid phase is found in a narrow region around
$\mu=-186$~K, and for lower values of the chemical potential
the density drops abruptly to zero. At low temperature, 
these results depict the system as a modulated self--bound 
superfluid which undergoes a first--order phase transition into 
the 1/6 commensurate crystal upon increasing the density.
A similar behavior is observed for the large cell (inset of 
Figure~\ref{fig_gf_nofmu_s}), apart from details in the somewhat
irregular increase of $N$ after the flat region, with a tendency
to form stripes in the small cell and domain walls in the large cell.
We believe that these differences are finite size effects caused
through steric constraints by the dimension and shape of the cells.  
We do not investigate this aspect further because our main interest
here is in the modulated liquid and the commensurate crystal phases.
For $T=0.25$~K we also simulate an
even larger cell with 11520 adsorption sites. Starting from the empty
cell, for $\mu=-186$~K we find the same liquid phase at areal density 
$\rho=0.050$~\AA$^{-2}$ as for the other cells, but for higher chemical
potential it takes too long to equilibrate a monocrystal at 1/6 covering.
However, a starting crystal configuration with 1920 atoms remains stable
even after very long runs for $\mu$ between -185 and -184~K.
\begin{figure}
\includegraphics[width=8cm]{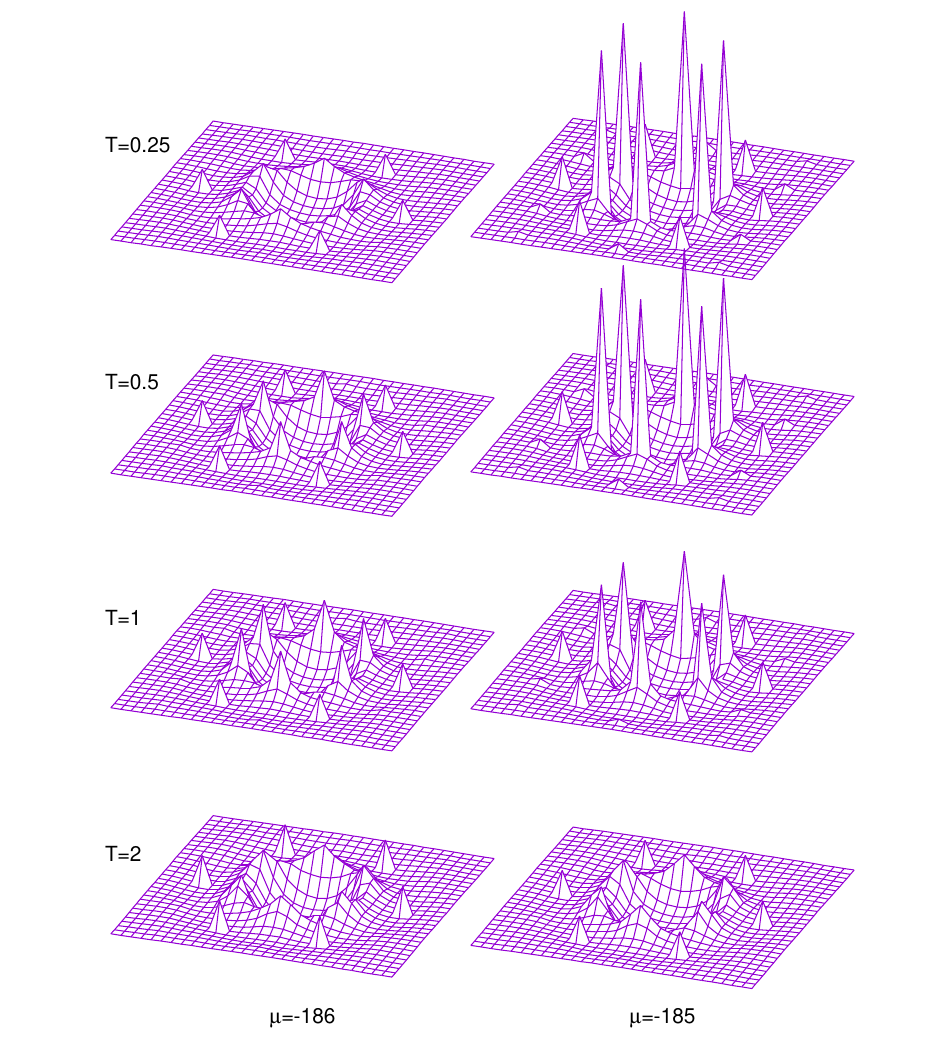}
\caption{
$^4$He adsorbed on GF:
two--dimensional structure factor calculated
in the small cell for various temperatures, as indicated in the body 
of the figure, 
at chemical potential $\mu=-186$~K (left panels) 
and $\mu=-185$~K (right panels).
The highest peaks are 6.2 tall.
}
\label{fig_sofk}
\end{figure}

These phase assignments are supported by structural data.
Figure~\ref{fig_sofk} shows the change in the
structure factor $S({\bf k})$ across the liquid--solid transition,
both in temperature and chemical potential. The six highest
peaks in each panel correspond to the first shell of reciprocal lattice
vectors, $|{\bf k}|=1.618$\AA$^{-1}$, of a triangular crystal which occupies
1/6 of the adsorption sites.

\begin{figure}
\includegraphics[width=8cm]{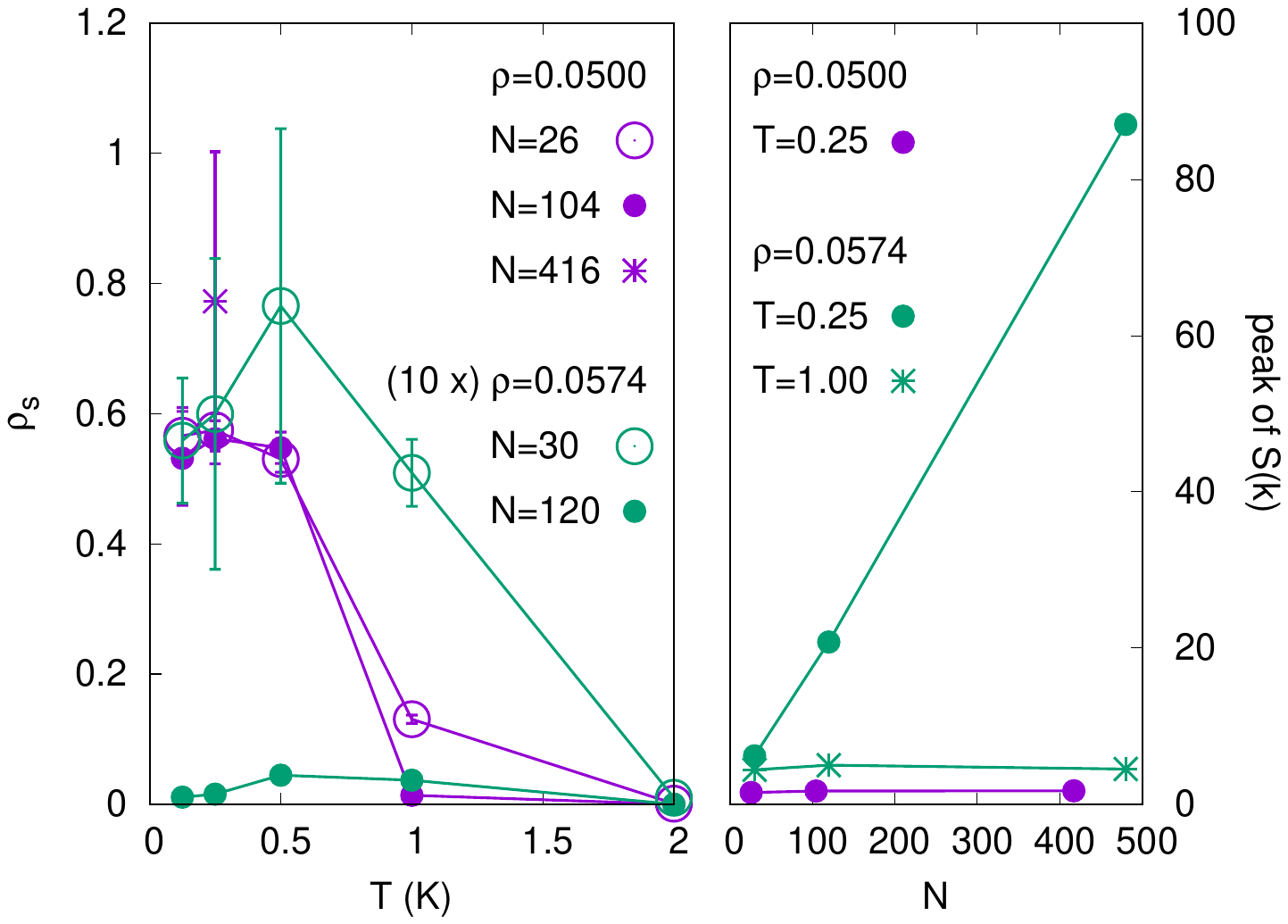}
\caption{
$^4$He adsorbed on GF:
dependence on the system size of the superfluid fraction and of the main peak
of $S(k)$ for two areal densities, $\rho\sim 0.050$~\AA$^{-2}$ (blue) 
and $\rho\sim 0.057$~\AA$^{-2}$ (green).
Left panel: $\rho_s$ as a function of $T$ for various numbers $N$ of particles;
for the higher density the data are multiplied by a factor of ten.
Right panel: main peak of $S(k)$ as a function of $N$ for $T=0.25$
and 1~K. 
}
\label{fig_size}
\end{figure}

For $\mu=-186$~K, $S({\bf k})$ features the characteristic 
ridge \cite{nava_2012} of a strongly modulated liquid.
At all temperatures six peaks are present at larger wavevectors with
$|{\bf k}|=2.802$~\AA$^{-1}$; these peaks arise because 
the fluid has a density modulation due to the substrate adsorption potential.
We note that the structure in the liquid phase is most pronounced at
$T=1$~K; for higher temperatures it is reduced by thermal
fluctuations, for lower ones by Bose exchanges \cite{ceperley_1995}.
The liquid has a finite superfluid fraction $\rho_s$ below a critical 
temperature located between $T=0.5$ and 1K.
This is shown in Figure~\ref{fig_size} (left panel, blue symbols), where 
$\rho_s$ is seen to exceed 50\% in a wide range of system sizes 
for $T\leq 0.5$~K,
while for $T=1$~K it has a much lower value which further drops toward zero
as the number of particles increases.
Even at the lowest $T$, $\rho_s$ is less than 100\% as it is expected on
general grounds for a nonuniform superfluid \cite{Leggett}.

For $\mu=-185$~K, at low temperature the system is in the commensurate 
crystal phase: 
the main peaks of $S({\bf k}$ soar to a value of 6.2 (Figure~\ref{fig_sofk}), 
which further increases linearly with the 
number of particles, as shown for $T=0.25$~K
by the green circles in the right panel of Figure~\ref{fig_size}.
On the other hand for $T=1$ the peak is still rather high 
(Figure~\ref{fig_sofk}), but it does not increase further for 
larger systems (green asterisks in the right panel of 
Figure~\ref{fig_size}). This indicates a melting
temperature between 0.5 and 1~K.
In addition to the main peaks, one notices additional weak peaks at 
larger $k$, related to the periodicity of the adsorption potential, 
as found in the superfluid state.
\begin{figure}[h]
\includegraphics[width=8cm]{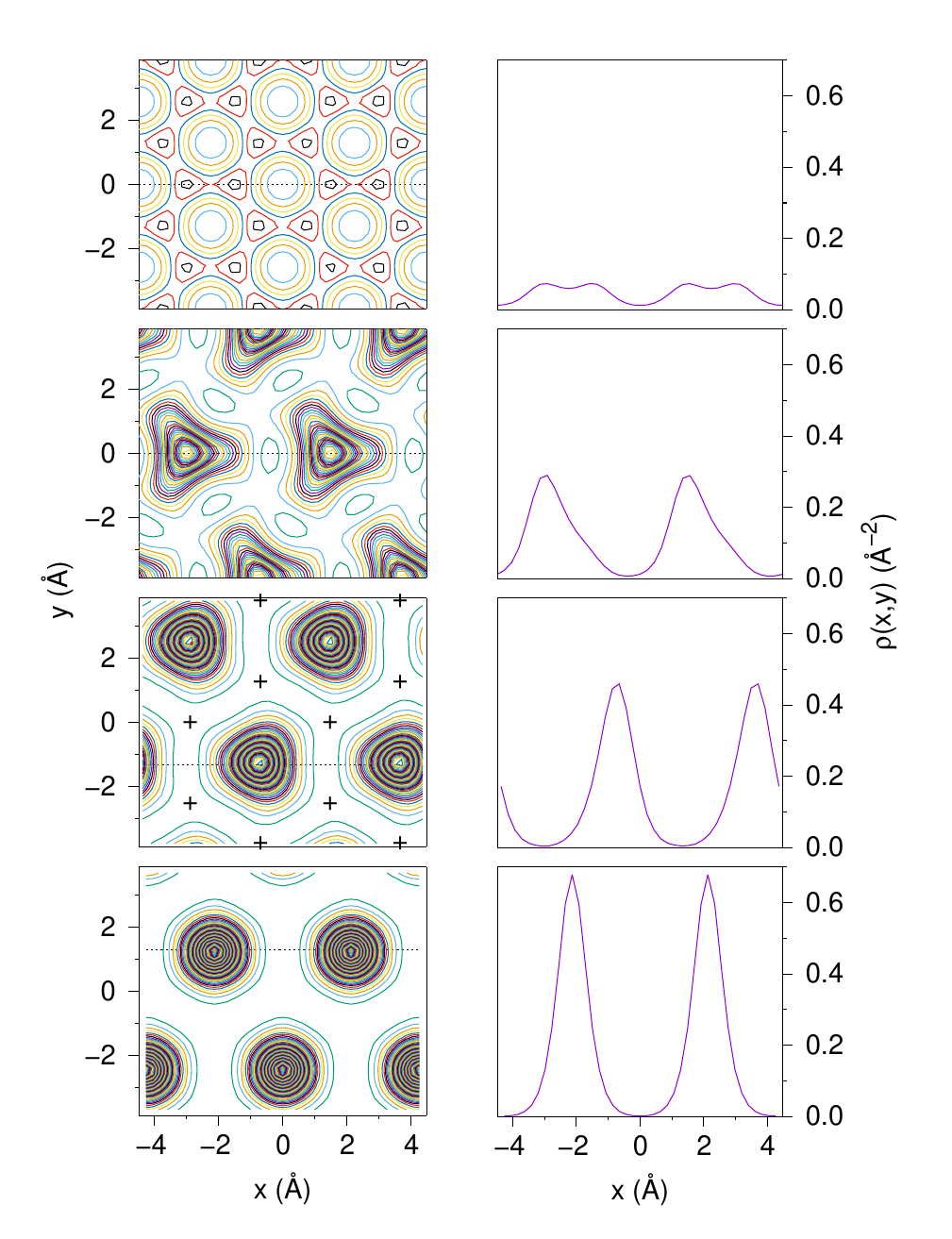}
\caption{
One-body areal density profile $\rho(x,y)$ of $^4$He adsorbed on various substrates at $T=0.25$~K.
Left panels: from top to bottom, superfluid at equilibrioum density on GF;
commensurate solid on GF; commensurate solid on hBN; commensurate solid on
G in the presence of an external electric field of 1~V/\AA. The contour 
level increment is 0.01,
and the lowest level is 0.02 in the top panel and 0.01 in the other ones. 
For GF, the adsorption sites correspond to the maxima (black contour lines) 
in the top left panel; for hBN, the (unoccupied) adsorption sites are 
marked with plus signs in the third left panel.
Right panels: $\rho(x,y)$
along the dotted line shown in the corresponding left panel.
}
\label{fig_rhoxy}
\end{figure}

The commensurate crystal is not supersolid: the small but finite
superfluid fraction calculated in this phase with the small cell drops 
to zero as the system size increases, even at the lowest temperatures 
considered (green symbols in the left panel of Figure~\ref{fig_size}).

Further details on the areal density profile $\rho(x,y)$ in the 
superfluid and the commensurate solid phases are given in Fig.~\ref{fig_rhoxy}. 
In the liquid phase (top panels) the $^4$He atoms distribute rather
uniformly along the potential valleys connecting the adsorption sites.
In the solid phase (second top panels) $\rho(x,y)$ features strongly 
anisotropic peaks centered on an adsorption site with shoulders on its
its three nearest neighbours and faint tails on its six second nearest 
neighbours. Such peaks are broader and lower than for commensurate
crystal phases of $^4$He on other substrates such as hBN or G.
The local density is spatially very anisotropic, the ratio between 
the largest and the lowest local density is about 40.

\begin{figure}[h]
\includegraphics[width=8cm]{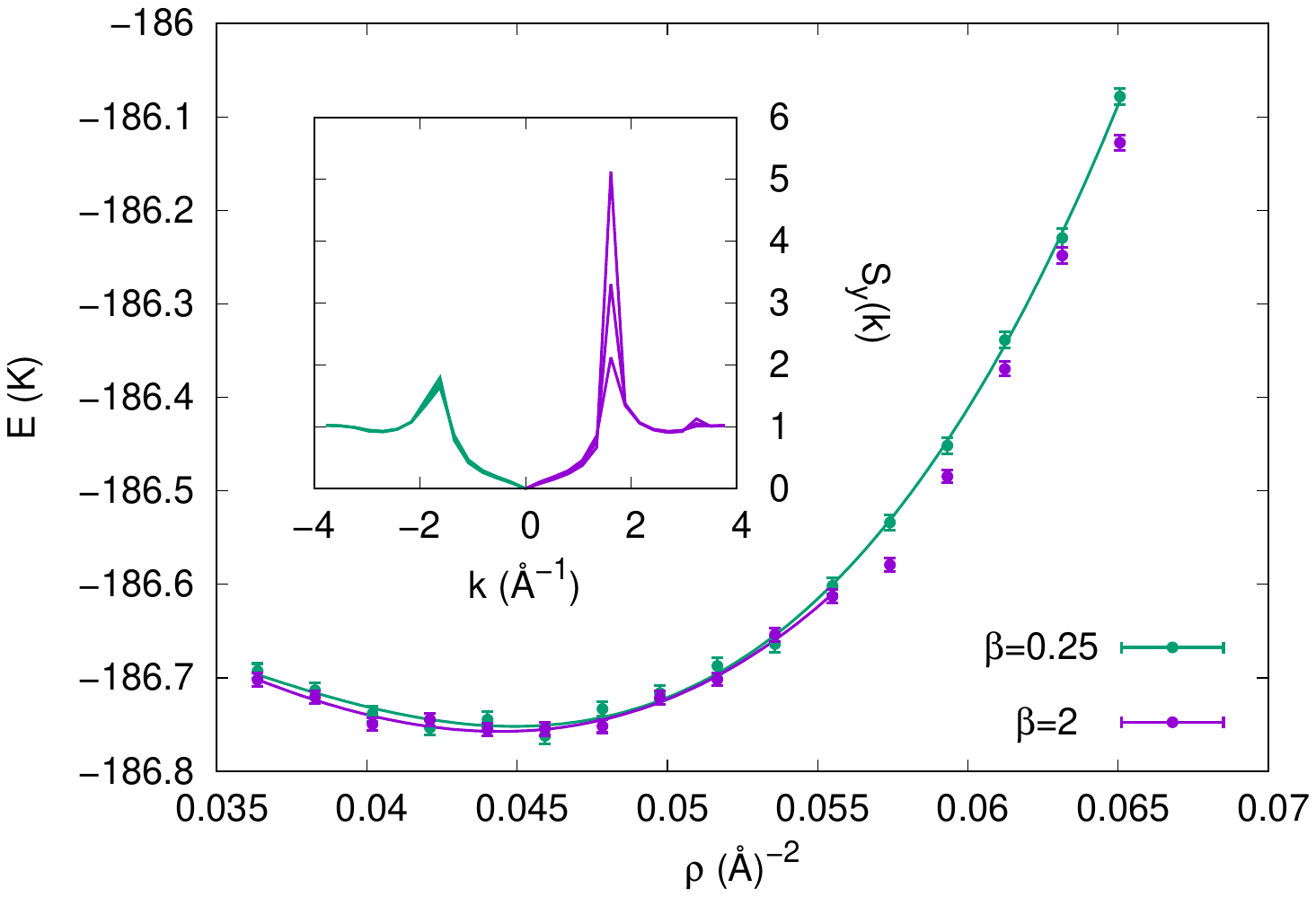}
\caption{
$^4$He adsorbed on GF:
energy per particle as a function of the areal density $\rho$ 
at $T=0$ calculated in the
small cell with projection time $\beta=2$~K$^{-1}$ (blue).
The line is a polynomial fit, restricted to the data in the liquid phase.
For areal density $\rho$=0.0574~\AA$^{-2}$, the system is in the
commensurate solid phase, in agreement with the finite 
temperature calculations.
The inset shows the structure factor along the $y$ axis of reciprocal 
space for $\rho=0.0536$, 0.0555 and 0.0574~\AA$^{-2}$, in order
of increasing peak heigth. For comparison, we include the energy and
the structure factor obtained with $\beta=0.25$~K$^{-1}$ (green).
}
\label{fig_eofn_GF}
\end{figure}

For $T=0$, in agreement with the low temperature results as expected,
we find a transition between a modulated liquid and a commensurate solid.
The energy per particle calculated for the small cell using the VPI method
with a projection time $\beta=2$~K$^{-1}$ is shown in 
Figure~\ref{fig_eofn_GF} with blue points.
For $\rho<0.0547$~\AA$^{-2}$ we can fit the Monte Carlo data with a
cubic polynomial, yielding an equilibrium density 
$\rho_0=0.044\pm 0.001$~\AA$^{-2}$. For larger densities, the energy
suddenly drops slightly below the fitted curve. In particular, for 
coverage 1/6, 
the structure factor is very similar (with minor quantitative differences 
discussed in the Appendix) to that shown in the top right panel of 
Figure~\ref{fig_sofk}, representative of the commensurate solid.
The estimated coexistence region extends between $0.0535\pm 0.005$
and 0.0574~\AA$^{-2}$.

We note that a large projection time is required to converge to the 
solid solution starting from the trial function of Eq.~(\ref{eq:psi}),
which represents a liquid. As shown by the green symbols and curves
in Figure~\ref{fig_eofn_GF}, a value of $\beta$ similar to that
employed in Ref.~\onlinecite{nava_2012} is perfectly adequate for
the liquid phase, but not sufficient for coverages 1/6 and higher.

\subsection{hBN}

hBN has one adsorption site per substrate unit cell, located at the hollow
point $H$ (see Figs.~\ref{fig:sites_gf} and \ref{fig_potentials}). 
In the $xy$ plane,
this corresponds to a minimum of the He-hBN potential, surrounded by 
saddle points at the $TB$ and maxima at the $TN$ sites. Therefore the 
corrugation of the absorption potential on hBN around an
adsorption site differs from that on G and on graphite: 
the angular periodicity is 120$^\circ$ for hBN and not 60$^\circ$ as for
G.
\begin{figure}[h]
\includegraphics[width=8cm]{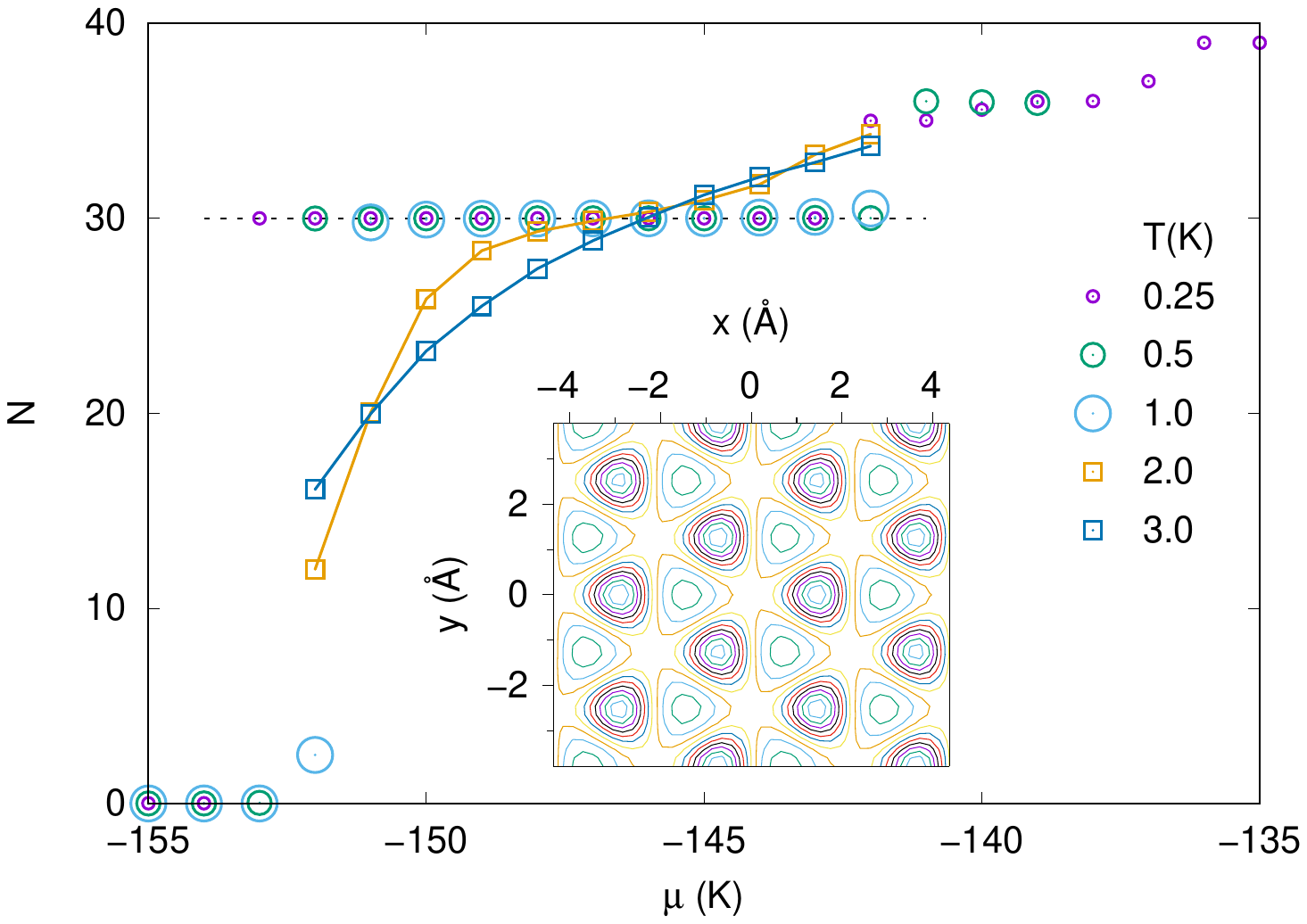}
\caption{
$^4$He adsorbed on hBN: number of atoms as a function of the chemical 
potential calculated for various temperatures, as indicated, 
for the small cell. We found similar results for the large cell at
$T=0.25$~K. The dashed line indicates the number of particles at
$\rho=0.0606$~\AA$^{-2}$, corresponding to 1/3 of the adsorption sites.
Inset: contour plot of the density profile in the normal fluid phase
at $T=3$~K, $\rho=0.0606$~\AA$^{-2}$; the contour level increment and
the lowest level are both 0.01.
\label{fig_hbn_nofmu_s}
}
\end{figure}

For sufficiently high temperature a monolayer of adsorbed $^4$He forms
a normal fluid phase. The density profile is directly shaped by the 
mentioned features of the corrugation potential: there is a peak 
centered on every $H$ site and elongated towards the three nearest $TB$ 
points, as shown in the inset of Fig.~\ref{fig_hbn_nofmu_s} for $T=3$~K.
\begin{figure}[h]
\includegraphics[width=9cm]{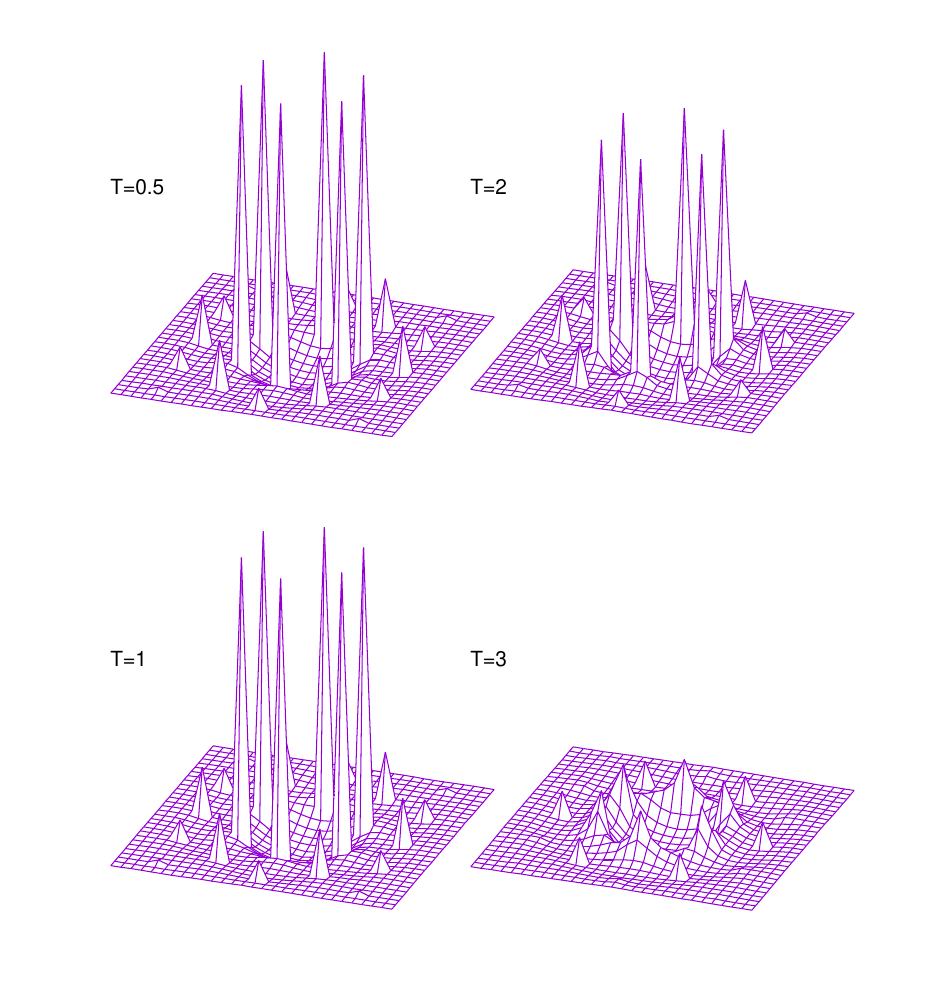}
\caption{
$^4$He adsorbed on hBN:
two--dimensional structure factor calculated
in the small cell for various temperatures, as indicated in the body 
of the figure, at $\rho=0.0606$~\AA$^{-2}$ ($\mu=-146$~K, see Fig.~\ref{fig_hbn_nofmu_s}). 
The highest peaks are 10.4 tall.
}
\label{fig_hbn_sofk}
\end{figure}

The dependence of the number of $^4$He atoms on the chemical potential,
obtained with PIMC simulations in the small cell for several temperatures,
is shown in Fig.~\ref{fig_hbn_nofmu_s}. 
For $T<2$~K, we find that the average areal density stays constant 
over a wide range of $\mu$, with a value $\rho=0.0606$~\AA$^{-2}$.
This corresponds to a commensurate solid, with $^4$He atoms occupying
1/3 of the adsorption sites. The analysis of the structure factor across
the melting temperature (which turns out to be about 2~K, 
Fig.~\ref{fig_hbn_sofk}) and its
dependence on the system size (similar to that shown in Fig.~\ref{fig_size} 
for GF) support this assignment. 
\begin{figure}[h]
\includegraphics[width=8cm]{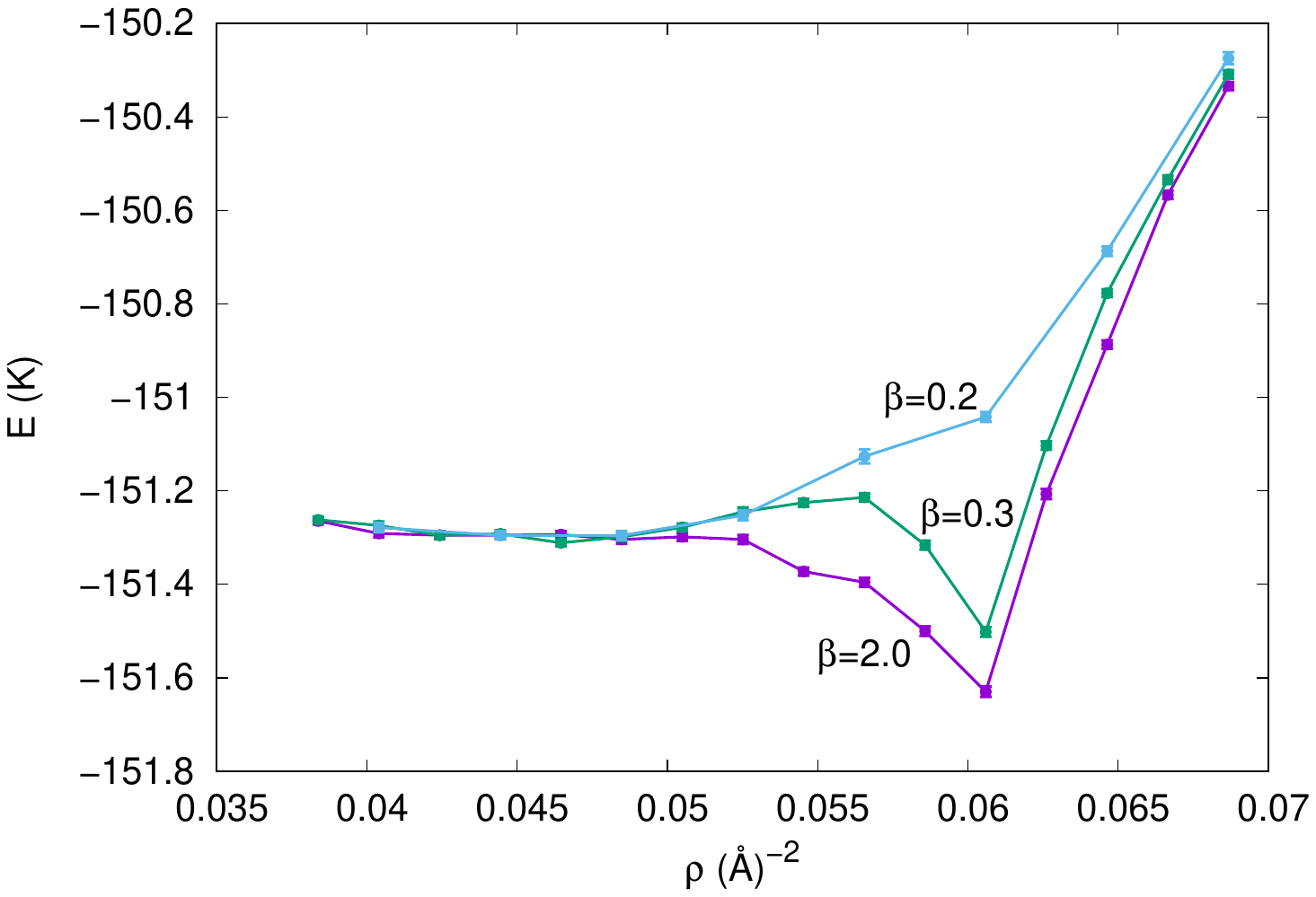}
\caption{
$^4$He adsorbed on hBN: energy per particle as a function of the
areal density $\rho$ calculated in the small cell with VPI simulations.
In order to illustrate the convergence toward the unbiased results for
projection time $\beta\to\infty$, we report energies calculated with
$\beta$=0.2, 0.3 and 2.0~K$^{-1}$.
\label{fig_eofn_hbn}
}
\end{figure}

From the heigth of the peaks of $S({\bf k})$ and from the density profile, 
plotted in the third--row panels of Fig.~\ref{fig_rhoxy},
we see that the commensurate solid on hBN is significanlty more localized 
than on GF. Analogously to the normal fluid phase at higher temperature,
the density peaks in the solid are elongated in the direction of the
three closest saddle points of the adsorption potential.

A more striking difference with the GF substrate is that,
in PIMC simulations carried out at low temperature ($T<0.5$~K), 
the system jumps directly from an empty cell to a commensurate 
crystal upon increasing the chemical potential. In particular, 
no (super)fluid phase should appear in the ground state of the 
system. 

This is confirmed by VPI calculations at $T=0$, which 
show that the commensurate solid is by far the lowest--energy 
state of the system (Fig.~\ref{fig_eofn_hbn}) and a fluid state is present
at a lower density as a metastable state. 
Just like in the case of GF, we note that this VPI result,
matching the findings of PIMC calculations at low temperature,
requires larger projection times than used in 
Ref.~\onlinecite{silvestrelli_2019}.

\subsection{Effect of an external electric field}
We have performed PIMC simulations of $^4$He at $T=0.25$~K in the 
presence of an external electric field $E=1$~V/\AA\ on hBN and 
$E=0.6$~V/\AA\ on G, 
the two substrates with the larger dependence on $E$ of the 
adsorption potential corrugation (see Table \ref{tab:table1}). 
 \begin{figure}[h]
\includegraphics[width=8cm]{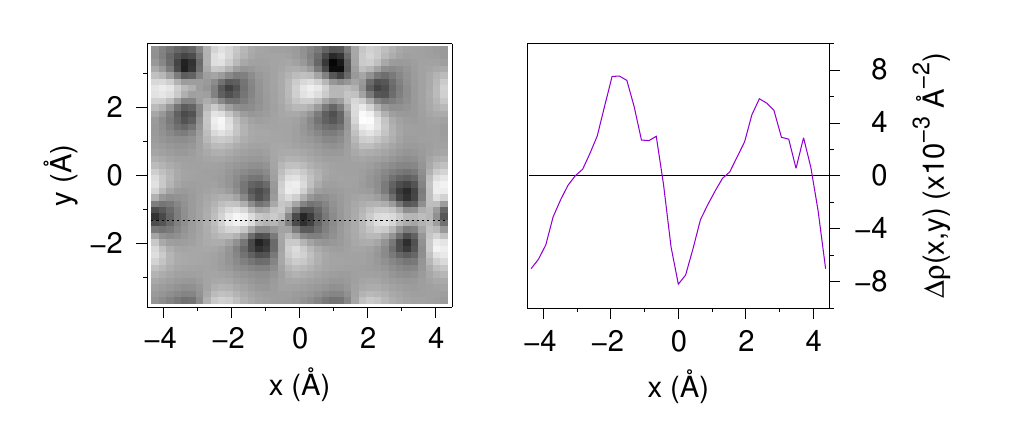}
\caption{
$^4$He adsorbed on hBN: variation of the density profile $\Delta\rho(x,y)$
induced by the external electric field. Left panel, greyscale map ranging
from -0.08 (black) to 0.08~\AA$^{-2}$ (white); each of the pictured objects 
with three negative and three positive lobes is centered on a density peak 
of the commensurate solid (see Fig.~\ref{fig_rhoxy}, third left panel).
Right panel, variation $\Delta\rho(x,y)$ along the dotted line shown in 
the left panel; the two structures displayed are not identical because of
statistical noise (note the small scale of the plot).
\label{fig_e}
}
\end{figure}

The external field tends to delocalize the adsorbed atoms: for hBN, the
range of chemical potential where  $^4$He is stuck in the commensurate solid
phase shrinks by $\sim3$~K; furthermore the peaks in the density profile get
further elongated towards the nearest saddle points of the He--substrate
potential. This can be seen by comparing the variation $\Delta\rho(x,y)$ of
the density profile, shown
in Fig.~\ref{fig_e}, with $\rho(x,y)$ itself (Fig.~\ref{fig_rhoxy}, third
left panel): the directions of the positive (white) lobes of
$\Delta\rho$ match those of the triangular stretching of the peaks.

However the effect is small (in fact, not visible on the scale of 
Fig.~\ref{fig_rhoxy}), and for both hBN and G we find that the commensurate 
solid remains stable, with no superfluid phases at lower densities. 

The density profile of the commensurate solid on G with external field 
is shown in Fig.~\ref{fig_rhoxy}; for a study of $^4$He adsorbed on G
without external field see Ref.~\onlinecite{kwon}).

\section{Summary and conclusion}
\label{sec_summary}
In this paper we have revisited the adsorption at low coverage of 
$^4$He on fluorographene and on hexagonal boron nitride in the 
sub monolayer regime and we have also studied such adsorption in presence 
of an electric field for GF and hBN and also for graphene.
The motivation was the search of a superfluid state of monolayer 
$^4$He adsorbed on very regular substrates, as can be experimentally 
obtained with the above-mentioned substances, because such
superfluid state should be characterized by a strong spatial anisotropy, 
a regime not yet explored with $^4$He. For instance, in this regime rotons 
should be anisotropic with an energy depending on the direction of the 
wave vector \cite{Nav13}
and a vortex excitation should not be translationally 
invariant, but it should have preferential sites for the location of the 
vortex core \cite{Gal2020}.

We have developed new adsorption potentials for such substrates with 
{\it ab initio} methods and we have studied the properties of $^4$He at finite 
temperature with PIMC and at $T=0$~K with VPI. In the case of GF we confirm 
previous results \cite{nava_2012,silvestrelli_2019}
that sub monolayer $^4$He in its ground state 
and at low $T$ is a low-coverage self-bound superfluid. A commensurate state 
is present at higher coverage in which one sixth of the adsorption sites 
are occupied. This commensurate state was not detected in the previous 
studies. In the present work we have not studied the system at higher 
coverage. The superfluid fraction at the lowest $T$ is about 55\% and 
not 100\%, a depletion expected even at $T=0$~K 
for a non-uniform superfluid \cite{Leggett}.
We estimate a transition temperature to the normal state
between 0.5 and 1.0~K. We have 
characterized the structural properties in direct and in reciprocal 
space confirming the extremely large spatial anisotropy of the superfluid. 
The superfluid fills the bonds of a honeycomb lattice of the adsorption
sites with a ratio of about 6 between the largest and 
the smallest areal density:
it is like the superfluid were moving in a 
multiconnected space. The commensurate state at coverage 1/6 of the 
adsorption sites is a triangular lattice similar to the commensurate 
state of $^4$He on graphite at coverage 1/3 but special features characterize 
the commensurate state on GF. In fact, even at the lowest $T$ a $^4$He 
atom is not constrained to remain at a single adsorption site but 
there is a sizeable probability of occupation of the neighboring sites 
giving rise to a three-lobed density distribution. 
It will be of interest to determine if such behavior, to our knowlegde
unique among the ordered adsorbates on regular substrates,
is modifying the order-disorder phase transition compared to that 
of $^4$He on graphite \cite{reatto_2013}.
Other features of this ordered state of $^4$He 
on GF are a mean square deviation from the equilibrium site more than double 
that of $^4$He on graphite and a  particularly low transition
temperature to a disordered state between 0.5 and 1~K, less than one third 
of the order-disorder transition of $^4$He on graphite. 
This commensurate state is not supersolid.

We find that the lowest energy state of $^4$He on hBN at $T=0$~K and at low $T$ 
is a commensurate triangular solid in which one third of the adsorption sites 
are occupied, a state isomorphous to that of $^4$He on graphite and graphene. 
The temperature of the order-disorder transition is about 2 K. The present 
result modifies the conclusion of a previous study \cite{silvestrelli_2019}
in which it was 
found that the ground state was a superfluid.  We have verified
that this discrepancy is 
due to the use of a too short 
projection time in the earlier $T=0$~K study. At coverage below this 
commensurate state a fluid state is present as a metastable one. 
We have shown that an electric field up to 1.0 V/\AA\ tends to delocalize the 
adsorbed atoms, but the effect is small and the electric field does not 
modify the phase behavior of the system. We find that also in the case of 
G the ground state of sub monolayer $^4$He remains an ordered state 
commensurate with the adsorption sites at coverage 1/3. 
\begin{figure}[h]
\includegraphics[width=4.5cm,angle=90]{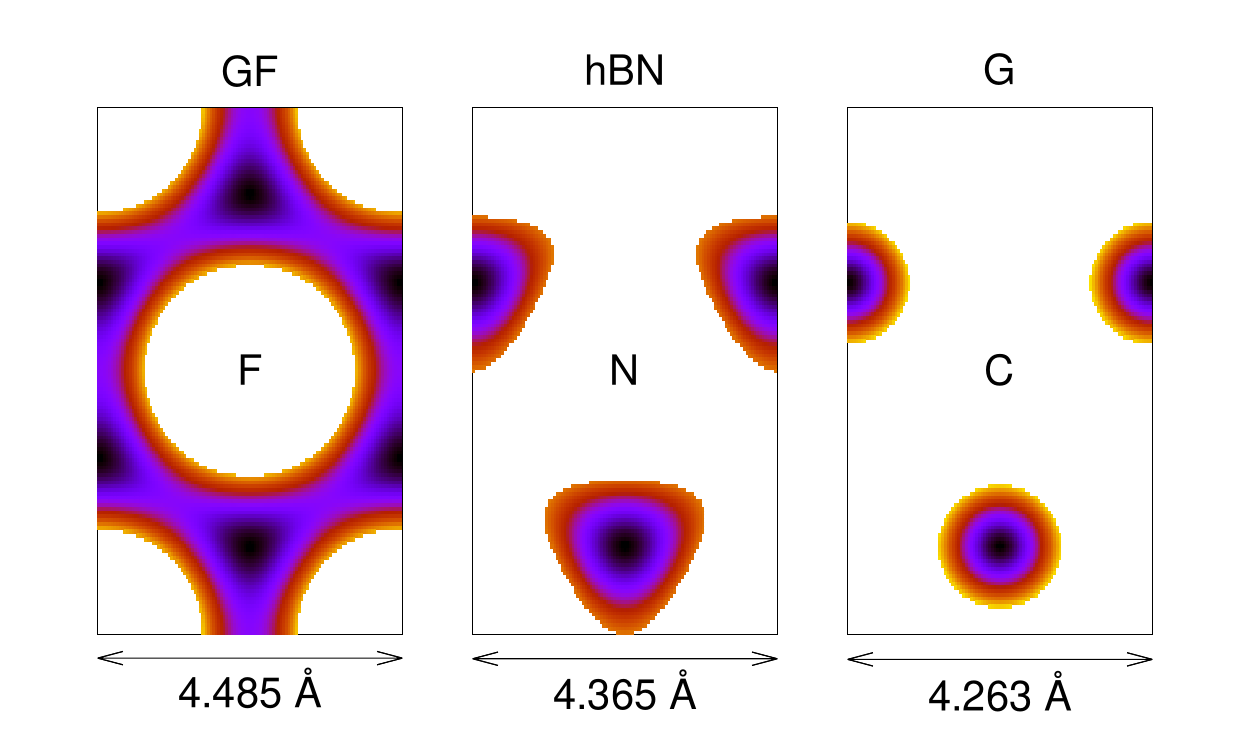} 
\caption{
The $x,y$ classically allowed regions for a $^4$He atom on GF, hBN and G
at a distance $d$ from the substrate corresponding to the minimum value 
$V_{\rm min}$ of the adsorption potential.
The color maps represent the adsorption potential $V(x,y)-V_{\rm min}$
from zero to the kinetic energy of a single $^4$He atom on the substrate
calculated with PIMC at $T=0.25$~K (i.e. 27.0, 22.6 and 28.4~K for
GF, hBN and G respectively). The color scale ranges from 0 (black) 
to 28.4~K (yellow). The plots are centered on the less favored site
(F atom for GF, N atom for hBN and C atom for G).
In the superfluid phase of $^4$He on GF (top panels of Fig.~\ref{fig_rhoxy}), 
the adsorbate atoms cover
rather uniformly the interconnected regions of low potential energy.
\label{fig_sites}
} 
\end{figure}

In conclusion, 
on present theoretical evidence the only regular substrate that supports 
a superfluid ground state is GF. Possibly this is also true 
for graphane \cite{nava_2012},
a compound isomorphous to GF with the fluorine atoms replaced by the hydrogen
ones, but experimentally it is difficult to produce regular substrate with 
full stoichiometry and we have not investigated this system with 
{\it ab initio} methods. 
The different nature of the lowest energy state of $^4$He on GF 
compared to that on hBN, G and graphite is mainly due to the 
number of adsorption sites per surface unit cell, twice as many
for GF as for the other substrates. As seen in Fig.~\ref{fig_sites}
this entails a larger, interconnected region of favorable, 
low potential energy available to the Helium atoms. 
The size and shape of such region stabilize
the superfluid phase and induce the three-lobed distortion of 
the peaks in the commensurate solid.


Other commensurate states might be present at larger coverage between the 
density range of the present study and the promotion coverage to the second 
layer, this is an interesting topic for future studies. Another interesting 
development is the study of the adsorption of the fermionic $^3$He on GF and 
on hBN. Due to the smaller mass of $^3$He it is very unlikely that the ground 
state of $^3$He on GF is a commensurate ordered state so that the low coverage 
state should be a liquid or an unbounded gas. Some evidence for a liquid 
state was found in the earlier study \cite{nava_2012}
with the semiempirical adsorption 
potential so further study of the fermionic system with the new {\it ab initio}
adsorption potential is of great interest because, in any case, $^3$He on GF
opens the possibility of studying a strongly interacting Fermi system with 
large spatial inhomogeneity. We have shown that the most stable state of 
$^4$He on hBN is ordered and a fluid state is only a metastable one. 
The smaller mass of $^3$He might alter the balance between these two states. 
In fact, we have performed PIMC computations for bosonic m=3 and indeed 
we find that with this mass the fluid state if the stable one. Of course, 
the Fermi statistics might alter this result, and this is left for future 
study. On this issue it will be important to perform new experiments on 
$^3$He on hBN to verify the earlier measurements \cite{crane_2000} 
that gave some evidence for a localized commensurate state.


\section*{Acknowledgments}
SM acknowledges support from the European Centre of Excellence in Exascale 
Computing TREX, funded by the European Union’s Horizon 2020 - Research 
and Innovation program - under grant agreement no. 952165.
P. L. S. acknowledges funding from Fondazione Cariparo, 
Progetti di Eccellenza 2017, relative to the project: 
``Engineering van der Waals Interactions: 
Innovative paradigm for the control of Nanoscale Phenomena''.

\section*{Appendix}
\begin{figure}[h]
\includegraphics[width=8cm]{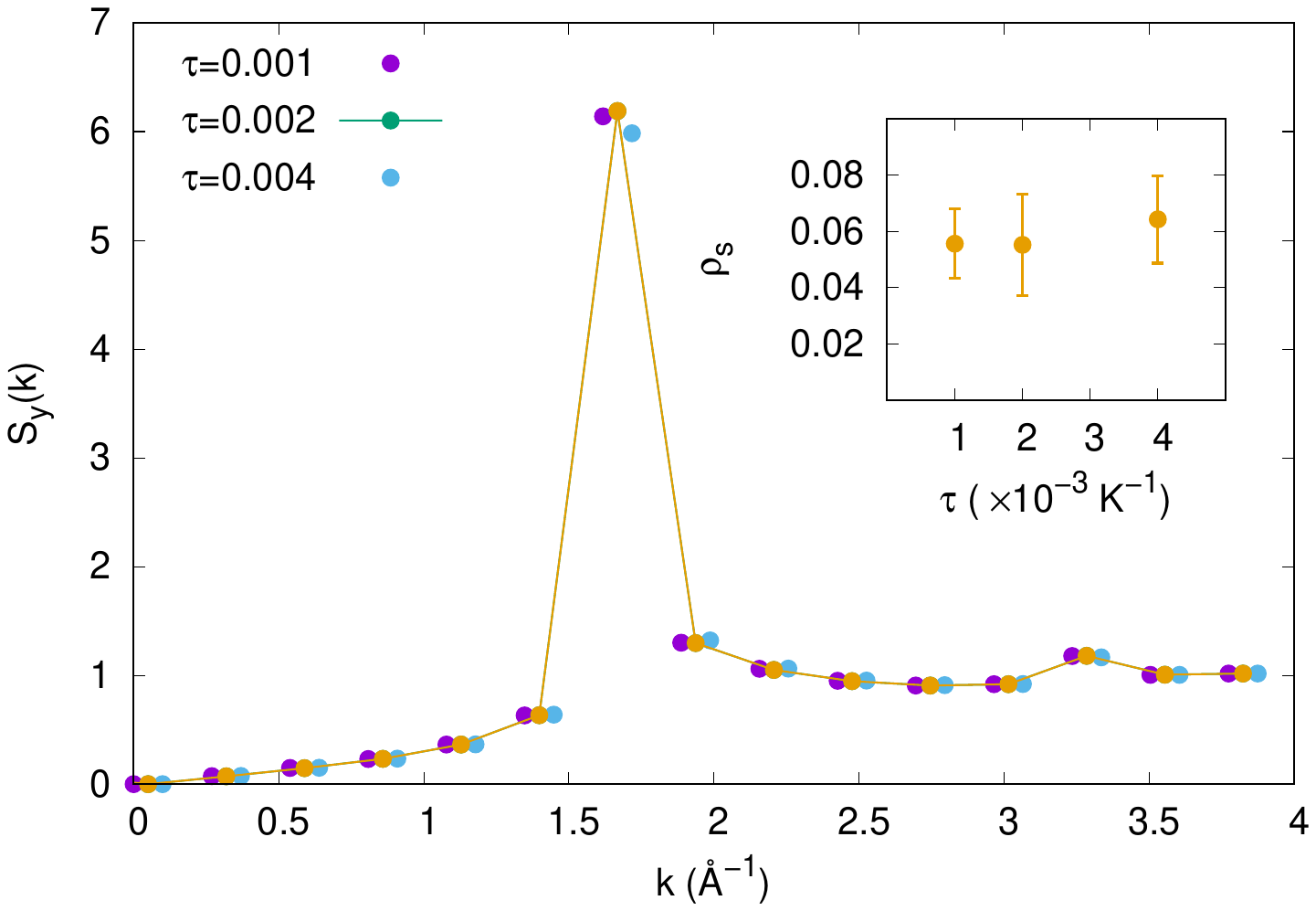}
\caption{
The structure factor along the $y$ axis of reciprocal space
for $N=30$ $^4$He atoms adsorbed on GF 
in the 1/6 commensurate solid phase, calculated with PIMC at $T=0.25$~K.
Data for different values of $\tau$ are slightly offset in $k$ for clarity.
The inset shows the superfluid fraction as a function of $\tau$.
}
\label{fig_sofkx}
\end{figure}

The PIMC simulations \cite{ceperley_1995,boninsegni_2006} give 
unbiased thermal averages for $\tau\to 0$. We use the primitive action 
with $\tau=0.002$~K$^{-1}$ which entails a negligible error in the
quantities of main interest here, namely the 
superfluid fraction $\rho_{\rm s}$, calculated using the so--called
winding number estimator,\cite{ceperley_1995}
and the static structure factor $S({\bf k})$,
calculated as the average of
$\sum_{ij}\exp[-i{\bf k}\cdot({\bf r}_i-{\bf r}_j)]\big/N$ over the
sampled configurations.
This is shown in Figure \ref{fig_sofkx} for $N=30$ $^4$He atoms adsorbed on GF
in the 1/6 commensurate solid phase (here the non--zero value of the
superfluid fraction is a finite--size effect).

\begin{figure}
\includegraphics[width=8cm]{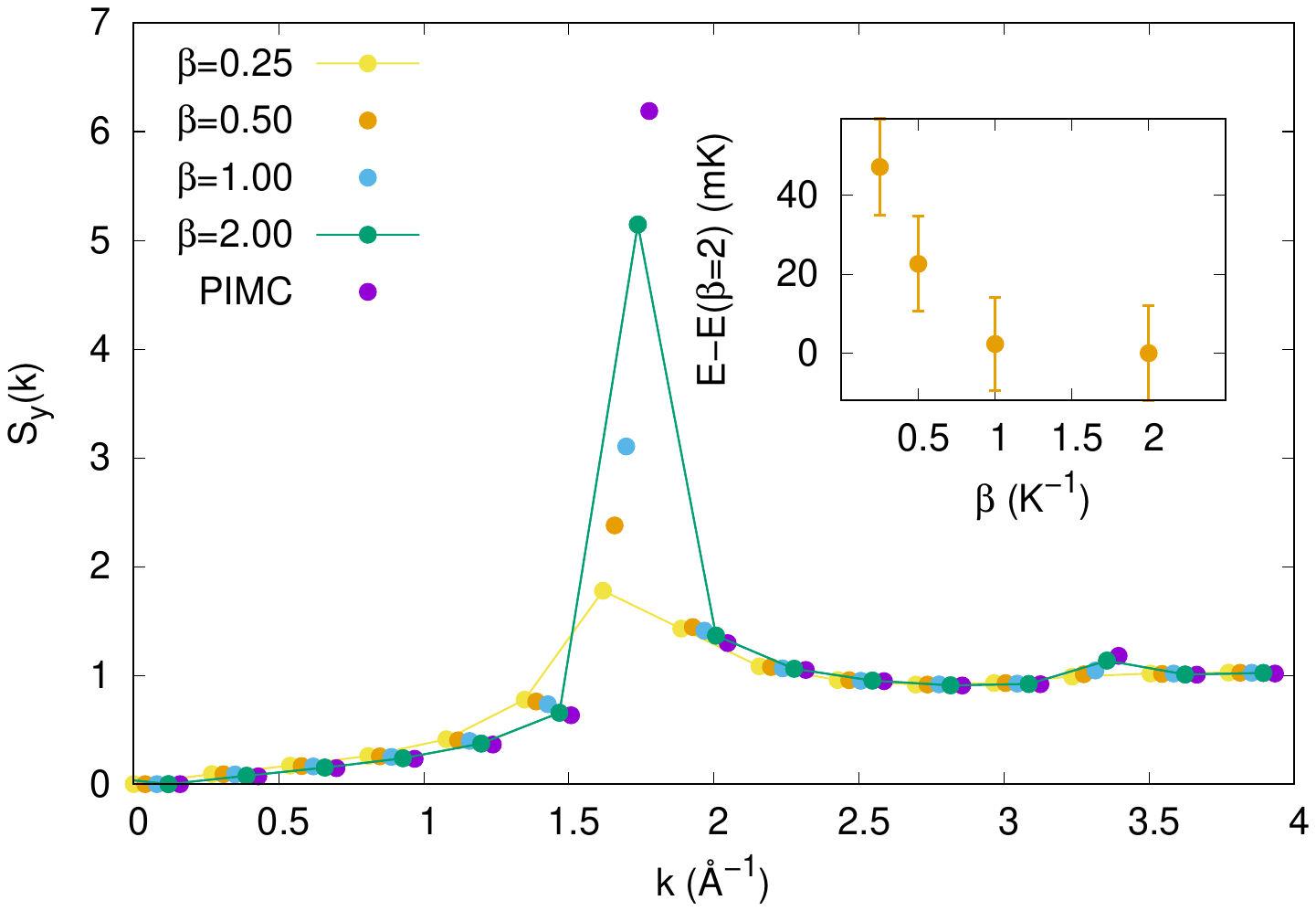}
\caption{
The structure factor along the $y$ axis of reciprocal space
for $N=30$ $^4$He atoms adsorbed on GF at the density of
the 1/6 commensurate phase, calculated with VPI.
Data for different values of $\beta$ are slightly offset in $k$ for clarity.
The PIMC result for $T=0.25$~K is also shown for comparison.
The inset shows the energy per particle as a function of $\beta$, relative
to its value at $\beta=2$~K$^{-1}$.
}
\label{fig_sofkbeta}
\end{figure}

The VPI simulations \cite{ceperley_1995,sarsa_2000} give unbiased
ground state results for time step $\tau\to 0$ and projection time 
$\beta\to\infty$. We use the primitive action with the same value 
$\tau=0.002$~K$^{-1}$ as in PIMC, and $\beta=2$~K$^{-1}$. 
The convergence in $\beta$ is shown in Figure \ref{fig_sofkbeta} 
for the structure factor and the energy.
In the range of $\beta$ between 0.25 and 2 the energy changes 
merely by $\sim 50$ mK, but the structure factor develops large peaks
at the first reciprocal lattice vector of the commmensurate solid. 
The convergence in $\beta$ of the main peaks is not complete,
because their heigth should be closer to that
obtained with PIMC at the lowest temperatures (also shown
in Figure~\ref{fig_sofkbeta} for comparison). 
Nevertheless, a projection time of $\beta=2$~K is clearly sufficient 
to turn the liquid state represented by the trial function of 
Eq.~(\ref{eq:psi}) into a solid.

\end{document}